\documentstyle[12pt]{article}
\begin{document}
\hoffset = -0.5truecm
\voffset = -3truecm
$\left.\right.$\\
\vspace{1cm}
\begin{center}

{\Large\bf Higher-genus corrections to black-string solution}\\
\end{center}
\vspace{3.5cm}
\centerline{{\large\bf M.Z.Iofa*}}
\begin{center}
{\small Nuclear Physics Institute,  Moscow State University,}\\
{\small 119899 Moscow, Russia.}\\
\end{center}
\vspace{3cm}
\centerline{{\large\bf Abstract.}}
\vspace{0.25cm}
One-string-loop (torus topology) corrections to black-string backgrounds
corresponding to gauged ${SL(2,R)\times R}/R$ WZW model are calculated 
using $\beta$-function equations derived from string-loop-corrected
effective action. Loop-corrected backgrounds are used to calculate ADM 
mass of the black string.

\vspace{5cm}

*E-mail address: iofa@theory.npi.msu.su
\newpage
{\bf 1. Introduction}

Recently much attention have recieved solutions of the 
gauged Wess-Zumino-Witten (WZW) 
$G/H$-models which provide conformal field theories interpreted as
describing geometries of black holes, black strings, etc. 
\cite{wi,hh,gq,ph,nwi,cr,fl,bs1}. These solutions 
satisfy $O(\alpha')\quad \beta$-function equations 
\footnote{See, however, paper \cite{st} where backgrounds 
of ${SL(2,R)\times R}/R$ model were shown 
to satisfy $\beta $-function equations
in $O({\alpha'}^2)$ approximation.}, but there are also 
conformally exact results valid in all orders in $\alpha'$ 
\cite{bs1,dvv,ts1}. However, 
usually these solutions are discussed for conformal field
theories defined on manifolds of topology of the sphere, i.e. at the tree
level of string-loop expansion. Some time ago, in papers \cite{emn}, 
string-loop corrections to the tree-level solutions
were discussed for 2D black-hole solution of gauged 
$SL(2,R)/U(1)$ WZW model. It was noted that in bosonic string theory, 
as a result of regularization 
of divergent integrals over the moduli, there appear imaginary 
corrections to the lowest genus solutions. Possible imaginary 
corrections to the mass of black hole could be interpreted as a manifestation 
of quantum instability of solution (stable at the tree level). However, 
in 2D theories the question of modular divergences
is somewhat ambiguous because in this case modular divergences are 
absent. 2D theory can be considered as a limit from $D>2$-dimensional models
which can have modular divergences, but in this setting the problem requires
more careful analysis.

At present, a large variety of gauged WZW models was investigated which yield
solutions interpreted as backgrounds of string theories in dimensions 
$D\,\geq\,3$ . In this paper, starting from loop-corrected renormalized 
string effective action (EA), we calculate loop corrections to tree-level 
backgrounds for the gauged ${SL(2,R)\times R}/R$ WZW model \cite{hh}
which is the first one from the set of ${SL(2,R)\times R^N}/R$ models 
\cite{gq,bs1} associated with
$D = (N + 2)$-dimensional backgrounds. Asymptotics of backgrounds are 
used to calculate string-loop-corrected ADM mass of the black string.

After fixing in sect.2 some notations, in sect.3 we introduce basic
formulas of Tseytlin's approach to construction of string-loop-corrected 
EA. In sect.4 general expressions are applied to the  one-loop case,
i.e. for the torus topology. In sect.5 we calculate asymptotics of 
background solutions to $\beta$-function equations. In sect.6 these
asymptotics are used to calculate loop-corrected ADM black-string mass.
In sect.7 we discuss duality for the loop-corrected solutions of $\beta$-
equations. Sect.8 contains concluding remarks and discussion.

{\bf 2.}
   
  Gauged WZW models provide a natural framework for Lagrangian realization
of coset models and form a bridge between conformal field theories and 
$\sigma$-model description of strings propagating in nontrivial backgrounds
\cite{wi,bn}.
For the gauged $ {SL(2,R)\times R}/R $ WZW model, after setting the axial
gauge and integrating out nonpropagating fields, in the limit of large 
central extension parameter $k$, in the leading order in $1/k$, one obtains
the action \cite{hh} 
\begin{eqnarray}
\label{E1}
\lefteqn{\hspace{-10mm}I=\frac {k}{4\pi} \int d^2 z\,\left 
\{-\left(1-\frac{1+\lambda}{r}\right) 
\partial t \bar{\partial}t +\left (1-\frac{\lambda}{r}\right)
\partial x \bar{\partial}x +
\frac{1}{4}\frac{\partial r \bar{\partial} r}
{(r-\lambda)(r-1-\lambda)} + \right.}
\nonumber\\
&&\left.+\sqrt{\frac{\lambda}{1+\lambda}}(1-\frac{1+\lambda}{r})
\left(\partial x \bar{\partial}t -\bar{\partial} x \partial t\right)
+\frac{1}{k} \sqrt{h}R^{(2)}(h) \Phi (r)  \right \}. \,
\end{eqnarray}  
With identification $\frac{1}{k}\rightarrow\alpha'$, where $\alpha'$ is the
string constant in dimensionless units, the action (1) 
can be interpreted as the action for the closed string propagating in
3D space-time equipped with the metric 
\begin{equation}
\label{E2}
ds^2= G_{\mu\nu}dx^{\mu}dx^{\nu} =-\left (1-\frac{1+\lambda}{r}\right) dt^2 +
\left (1-\frac{\lambda}{r}\right)dx^2 +
\frac{1}{4}\frac{dr^2}{(r-\lambda)(r-1-\lambda)}\,
\end{equation}
antisymmetric tensor gauge field   
\begin{equation}
\label{E3}
B_{tx} = \sqrt{\frac{\lambda}{1+\lambda}} \left(1-\frac{1+\lambda}{r}\right)\,
\end{equation}
and dilaton
\begin{equation}
\label{E4}
\Phi =  \frac{1}{2}(a-\ln r).\,
\end{equation}
Backgrounds (2)-(4) are solutions of equations of motion derived from
the $O(\alpha')$ part of string effective action (EA) \cite{hh,gq,st}
\begin{equation}
\label{E5}
S = a_0 \int d^Dx\,\sqrt{|G|}\,
e^{-2\Phi} \,\left[\Lambda-\frac{\alpha'}{2}\left(\hat{R}+4D^2\Phi-
4(D\Phi)^2\right) + O(\alpha'^2)\right]\,
\end{equation}
for $D=3$. Here
$$ \Lambda=\frac{D-26}{3}; \qquad \hat{R}=R - \frac{H^2}{12}    $$
 Equations of motion following from (5) are equivalent to conditions 
of Weyl invariance of the theory with the $\sigma$-model action  
\begin{equation}
\label{E6}
I=\frac{1}{4\pi\alpha'}\int d^2z\sqrt{h}\,\left[(G_{\mu\nu}h^{ab}+ 
B_{\mu\nu}\frac{\varepsilon^{ab}}{\sqrt{h}}) \partial_a x^{\mu}\partial^a 
x^{\nu}+ \alpha' R^{(2)}\Phi(x) \right]
\end{equation}
where $h_{ab}$ is the world-sheet metric,
and can be symbolically written as \cite{cfmp,ts2}
\begin{equation}
\label{E7}
\bar{\beta}^i (\varphi^i)=0. \,
\qquad (\varphi^i=G_{\mu\nu}, B_{\mu\nu},\Phi)\,
\end{equation}

{\bf 3.}

  In closed bosonic string theory, the full string EA is obtained as the 
renormalized generating function of massless string amplitudes
$$ S(\varphi_R) = \hat{Z}_R (\varphi_R)=\hat{Z}(\varphi(\epsilon),\epsilon) $$
where the bare fields $\varphi(\epsilon)$ are expressed as perturbative
expansion in $\ln \epsilon$ with the coefficients depending on renormalized 
fields $\varphi_R$. Generating function $\hat{Z}(\varphi(\epsilon),\epsilon)$
is constructed as the sum of contributions from all the genera $\chi =2-n$
\begin{equation}
\label{E8}
\hat{Z}(\varphi(\epsilon),\epsilon) = \frac{\partial}{\partial \ln \epsilon}
\sum_{n=0}^{\infty}\bar{Z}_n (\varphi(\epsilon), \epsilon)\,
\end{equation}
where $n$ is the number of handles of the world-sheet surface \cite{ts4,ts5}.
 Here
\begin{equation}
\label{E9}
\bar{Z}_n = \int [d\mu (\tau, \epsilon)]_n \,Z_n\,
\end{equation}
and
$$Z_n = a_{n}\int D[x,h]'\sqrt{|\det G|} e^{-I}$$ 
is the partition function obtained by 
integration over all (functional) variables except for the moduli $\tau$ 
of the world-sheet metric $h$.

    The basic assumption about $\hat{Z}(\varphi(\epsilon),\epsilon)$ is that
it is perturbatively renormalizable both in $\alpha'$ and string-loop 
expansions \cite{ts4,ts5}. An important aspect of this property is that all 
divergences (modular and ultraviolet) are regularized in a universal way by 
the same cutoff parameter $\epsilon$. The derivative with respect to 
$\ln\epsilon$ takes care of the properly accounted divergent volume of
the M\"obius group.

   An explicit realization of such regularization is provided by Schottky
parametrization \cite{lo,al,ky,ts5} of the extended moduli space. 
In this parametrization,
a surface of genus $\chi=2-n$ is mapped on the complex plane $\bf{C}$ 
\footnote{More exactly, a surface is mapped on compactification of $\bf{C}$
i.e. on the 2-sphere \cite{ts5}.} with $n$
pairs of holes with the pairwise identified boundaries. On the complex plane 
$\bf{C}$ acts the group $SL(2,\bf{C})$. If the corresponding M\"obius 
symmetry is not fixed, then the volume of the group $SL(2,\bf{C})$ enters the 
amplitudes as the universal divergent factor \footnote{This is true for 
$n\geq 3$-point amplitudes.}. Fixing 3 complex parameters of the group
$SL(2,\bf{C})$, one reduces the number of independent moduli to $3n-3$.

   Divergences of the amplitudes can appear either if positions of several 
vertex operators (punctures) tend to each other or/and if the holes from the 
handles shrink to a point. In Schottky parametrization, all the divergences
are universally regularized by introducing the "minimal distance" $\epsilon$
which enters propagators as well as integration measure over the moduli.

   Partition function $Z_n$ has the following form \cite{ts4}
\begin{equation}
\label{E10}
Z_n=a_n e^{\frac{\chi\Lambda}{2} \ln\epsilon}\,\int d^Dx\,\sqrt {|G|}\,
e^{-\chi\Phi}\,\left [1+\alpha'\left(b_1^{(n)}\hat{R}
+ b_2^{(n)}D^2\Phi\right)+\ldots \right], \,
\end{equation}        
where
$$ b_1^{(n)} = \frac{\pi}{V}\int d^2z\sqrt {h}\,{\it G}(z,z')-
\pi\int d^2z\sqrt {h}\,\left[\nabla_a\nabla'^a {\it G}(z,z')\,
{\it G}(z,z')-\left (\nabla^a {\it G}(z,z')\right )^2 \right]_{z=z'}\,$$
\begin{equation}
\label{E11}
 b_2^{(n)}= -\frac{1}{4}\int d^2z\sqrt{h}\,R^{(2)}(h) {\it G}(z,z) \,
\end{equation}
$$ V=\int d^{2}z\sqrt {h}.  $$
Here ${\it G}(z,z')$ is the regularized propagator on the world sheet of
genus $\chi=2-n$ with a metric $h$ (regularized Green function of scalar 
Laplacian).

The coefficients $b_{1,2}^{(n)}$ contain logarithmically divergent parts 
which appear from the limit of coinciding arguments in the propagators
\begin{equation}
\label{E12}
b^{(n)}_{1}=\frac{1}{2}\ln\epsilon+\bar b^{(n)}_{1}  ;
 \quad b^{(n)}_{2}=(n-1)\ln\epsilon+\bar b^{(n)}_{2},\,
\end{equation}   
and are defined up to transformations of the finite parts
$\bar{b}_{1,2}^{(n)}$ under reparametrizations of the fields $\varphi^i $
\begin{equation}
\label{E13}
 G_{\mu\nu} \rightarrow G_{\mu\nu}+\alpha'(a_1 R_{\mu\nu}+a_2 G_{\mu\nu}R+
a_3 D_{\mu} D_{\nu}\Phi+a_4 G_{\mu\nu}D^2 \Phi+\ldots )
+\ldots,\, 
\end{equation}
$$\Phi \rightarrow \Phi +\alpha'b_1 R+\ldots,\quad
B_{\mu\nu} \rightarrow B_{\mu\nu}+\alpha'c_1 D_\lambda H_{\mu\nu}^\lambda +
\ldots $$
which do not change the massless sector of the (tree) string theory
$S$-matrix \cite{ts4,ts5}.
   Renormalized partition function $Z^R_n(\varphi_R)$ is obtained by 
substituting expressions for bare fields in terms of renormalized fields. 
In the leading order in $\alpha'\ln\epsilon$ one has
\begin{equation}
\label{E14}
\varphi^i = \varphi^i_R - \beta^i(\varphi_R)\ln \epsilon/\mu + \ldots. .  \,
\end{equation}

{\bf 4.}

 The tree-level (topology of sphere) generating functional 
$\bar{Z}_0$ contains
no integration over the moduli. At the one-string-loop level (topology of
torus), the "extended" moduli space is parametrized by three complex
parameters $\xi, \eta \; \mbox{and}\; k$. The measure on the "extended" moduli 
space is 
\begin{equation}
\label{E15}
 d\mu_{1}=\frac {d^{2}\xi d^{2}\eta}{{\left|\xi-\eta\right|}^{4}} [d^2k]. \,
 \end{equation}
 Here $|\xi-\eta|$ has the meaning of the distance between the centers of 
the holes from the handle on the complex plane $\bf C$ \footnote{To be
precise, the measure (14) can be used to calculate $N \geq 3$-point 
amplitudes. To calculate $N \leq 3$-point amplitudes and, in particular,
the vacuum functional $\bar{Z}_1$ some modifications are required \cite{ts5}
yielding the final expression (18).}.
 In parametrization $ k=e^{2\pi i\tau}$ ,
the measure $ [d^2 k] $ is given by
\begin{equation}
\label{E16}
    [d^{2} \tau]=\frac{d^2 \tau}{\tau_2^2}\left(
    {\tau_2}\,{|\eta(\tau)|}^4\right)^{-\frac {D-2}{2}},\,
\end{equation}
where $\eta(\tau)$ is the Dedekind $\eta$-function
   $\,\eta (\tau) = k^{\frac{1}{24}}\prod_{1}^{\infty}(1-k^m)$.
Note that ${ k\sim e^{-2\pi\tau_{2}}}$ as
$\,\tau_{2}\rightarrow\infty\,$.
Propagator on the complex plane $\bf C$ with two discs from the handle
removed is
\begin{equation}
\label{E17}
G(z_1,z_2)= -\frac{1}{4\pi} \ln\left\{\left ({|z_1 - z_2|}^2 +
\epsilon^2 \right)
\prod_{m=1}^{\infty}\frac{(1-\lambda{k}^m)
(1-{\lambda}^{-1}{k}^{m})}{{(1-{k}^{m})}^{2}}\right\}+
\frac{{(\ln |\lambda |)}^2}{2\pi\tau _2}\,,
\end{equation}
 where
$$\lambda =(z_{1}-\xi)(z_{2}-\eta){(z_{1}-\eta)}^{-1}{(z_{2}-\xi)}^{-1}.\,$$
Singularity at $\xi =\eta$ in the measure (\ref{E15}) is regularized by the 
same cutoff as in the propagator (\ref{E17}):
$ |\xi-\eta|^2 \rightarrow |\xi-\eta|^2 + \epsilon^2. $

Performing integrations in the formulas (\ref{E11}), one obtains the generating
functional $\bar{Z}_1$ in the form \cite{ts4,ts5}
\begin {equation}
\label{E18}
\bar{Z}_1 = a_{1}\int d^Dy\sqrt{|G|}\,\int[d^2\tau]\,\left[\ln\epsilon
\left(1 +
\frac{\alpha'}{2}\hat{R}\ln\epsilon +\alpha'(b^{(1)}_1\hat{R}+b^{(1)}_2 (D\Phi)^2 )  
\right) + O({\alpha'}^2) \right].\,
\end{equation}
Here the first logarithmic factor appears from integration over the moduli, the
second one is due to ultraviolet divergences in the propagators at the
coinciding arguments. All constant factors are 
included in $a_1$. \footnote {Substituting propagator (\ref{E17}) in expressions
(\ref{E11}), we obtain that $\bar{b}^{(1)}_{1} =\bar{b}^{(1)}_2=0$.
However, nonzero terms $\bar{b}^{(1)}_{1,2}$ can be generated by 
transformations (\ref{E13}).}

Collecting  tree-level and one-loop contributins, one has
\begin{eqnarray}
\label{E19}
\hat{Z}(\varphi(\epsilon), \epsilon)&=&a_0e^{\Lambda \ln \epsilon}
\int d^Dx\,\sqrt {|G|}\,
e^{-2\Phi}\left[\Lambda-\frac{\alpha'}{2}\left(\hat{R}+4(D\Phi)^2\right)+
O(\alpha'^2)\right]+ \\
&& a_1\int [d^2\tau]\,\int d^Dx\, \sqrt{|G|}\,\,
[1 + \alpha'\hat{R}\ln \epsilon+\alpha'(b^{(1)}_1 \hat{R}+b^{(1)}_2 (D\Phi)^2 )+ 
O({\alpha'}^2)] \nonumber 
\end{eqnarray}
Renormalized generating functional $\hat{Z}_R$ is obtained by substituting
expressions for bare fields in terms of renormalized ones and taking into
account additional terms from string-loop divergences (cf. with (\ref{E13}))
\cite{ts4,ts5}.
\begin{equation}
\label{E20}               
\varphi^i = \varphi^i_R - \beta^i(\varphi_R)\alpha'\ln \epsilon/\mu +
\delta\beta^i(\varphi_R)\alpha'\ln \epsilon/\mu + \ldots \,
\end{equation}
Additional terms $\delta\beta^i(\varphi_R)$ are obtained from the requirement
to cancel string-loop $\ln \epsilon$ terms in (\ref{E18}) and are equal to
\begin{equation}
\label{E21}
\delta\beta^i = \rho e^{2\Phi^R}\left(\frac{d}{4},\quad 
\frac{1}{2}g^R_{\mu\nu},\quad B^R_{\mu\nu} \right), \,
\end{equation}
where $\rho=2a_1 / a_0  \int [d^2 \tau]$.
Substituting (\ref{E20}) in (\ref{E19}), we obtain the expression for the 
renormalized generating functional $\hat{Z}_R $ which includes contributions 
from sphere and torus topologies
\begin{eqnarray}
\label{E22}
\hat{Z}_R&=&a_0 \int d^Dx\,\sqrt {|G|}
e^{-2\Phi}\,\left[\Lambda-
\frac{\alpha'}{2}\left(\hat{R}+4(D\Phi)^2\right)+\right.\nonumber\\
&&\left.\frac{\rho}{2}e^{2\Phi} \left(1+\alpha'\hat{R} \ln\mu 
+\alpha'(b^{(1)}_1\hat{R}+b^{(1)}_2 (D\Phi)^2 )\right)+\ldots\right]\,
\end{eqnarray}
(henceforth we omit the subscript $R$). It is seen that the formal effective 
parameter of string-loop expansion is $\rho e^{2\Phi} $. Using the freedom in
the choice of reparametrization of fields (\ref{E13}) the terms $\alpha'\hat{R} 
\ln \mu +\alpha'(b^{(1)}_1\hat{R}+b^{(1)}_2 (D\Phi)^2) $ can be set to zero.
Adding to the action (\ref{E22}) the total derivative $ 2D^2(e^{-2\Phi}) $ to 
have the same tree-level part of the action as in (\ref{E5}), one finally 
obtains the renormalized EA \begin{equation}
\label{E23}
S=a_0 \int d^Dx\,\sqrt {|G|}
e^{-2\Phi}\,\left[\Lambda-\frac{\alpha'}{2} \left(\hat{R}+4D^2\Phi-
4(D\Phi)^2 \right)+\frac{\rho}{2} e^{2\Phi} \right].\,
\end{equation}

{\bf 5.}

 Our next aim is to find string-loop corrections to tree-level solutions
of eqs. (6). Variation of EA (\ref{E22}) yields the following equations of motion:
\begin{eqnarray}
\label{E24}
\bar{\beta}^G_{\mu\nu} + \delta\bar{\beta}^G_{\mu\nu} & = & \hat{R}_{\mu\nu} +
2D_\mu D_\nu \Phi  +\frac{\rho }{2\alpha'} G_{\mu\nu}e^{2\Phi} = 0, \\
\label{E25}
\bar{\beta}^\Phi & = & \hat{R} - \frac{2\Lambda}{\alpha'} - 
4(D\Phi)^2 + 4(D^2 \Phi) = 0, \\
\label{E26}
\bar{\beta}^B_{\mu\nu}  & = &  D^{\lambda}\Phi H_{\lambda\mu\nu} = 0. 
 \end{eqnarray}

 Tree-level backgrounds (2)-(4) depend on a single parameter $r$ and
provide an example of the "rolling moduli" solution \cite{mu} to non-linear 
eqs. (\ref{E7}). In the following, having in view calculation of the mass of the
black string, we shall be interested in finding asymptotics of solutions
to nonlinear equations (\ref{E24})-(\ref{E26}) at $r \rightarrow \infty$. 
In this limit,
we can linearize the system (\ref{E24})-(\ref{E26}) and solve it explicitly.

Solving the tree-level eqs. (\ref{E24})-(\ref{E26}) (with $\rho=0$), one
obtains the "rolling" solution for the metric and dilaton (together with the 
corresponding solution for the antisymmetric tensor) of the form 
\begin{equation}
\label{E27}
ds^2 = -\frac{\sinh^2 \frac{\gamma z}{2}}{\lambda + \cosh^2 \frac{\gamma
z}{2}}dt^2 + \frac{\cosh^2 \frac{\gamma z}{2}}{\lambda + \cosh^2 \frac{\gamma
z}{2}}dx^2 + dz^2,
\end{equation}
$$\Phi=\frac{1}{2}(a-\ln\cosh^2 \frac{\gamma z}{2})$$
where $ \gamma^2 = \frac{2|\Lambda|}{\alpha'}$. Taking $\gamma^2 =4$ ($\alpha'$
in dimensionless units), and introducing new variable $r$ by the relation
$$r = \lambda + \cosh^2 z $$
one obtains the solution (\ref{E2})-(\ref{E4}) of the gauged WZW model.
The metric and dilaton (\ref{E27}) are asymptotic to the flat-space solution
$(\eta_{\mu\nu},\,\Phi^0)$, where $\Phi^0=\frac{1}{2}(a-\gamma |z|)$. 
Flat-space solution can be considered as the limiting form of the "rolling"
solution (\ref{E27})  as $\alpha' \rightarrow 0 \quad (\gamma \rightarrow
\infty)$ \footnote{Dilaton $\Phi^0$ is composed from two branches of
solutions to the flat-space dilaton equation $\frac{\gamma^2}{4}=
({\Phi^0}')^2$. In the following, this results in "constants" having the 
$\mbox{sgn}z$ factor. Note that we are interested only in asymptotic region
of large $|z|$ where, up to exponentially small terms, one can use a smooth
approximation of $\Phi^0$, for example, $\Phi$ from (\ref{E27}).}.
 
To solve loop-corrected equations, let us introduce new variable $z$ by the 
relation $r = \lambda + {\cosh}^2 \frac{\gamma z}{2} + f(\rho, z)$,
where the function $f$ is chosen so that in new variables the $zz$ component
of the metric is again equal to unity. Asymptotically as
$z\rightarrow\infty, \quad f(z)=O(\rho z^n e^{-\gamma |z|})$ with some $n$.
As in the case $\rho = 0$, we are looking for solution for the metric and 
dilaton  asymptotic to the flat-space solution.
Writing the metric and dilaton as 
$$ G_{\mu\nu}=\eta_{\mu\nu}+h_{\mu\nu}$$
$$\Phi=\Phi^0+\varphi,\,$$ 
where $h_{\mu\nu}$ and $\varphi$ are of order
$O(\rho z^n e^{-\gamma |z|})$ and linearizing the equations about the
flat-space solution, we obtain
\begin{equation}
\label{E28}
h_{tt}'' -2{\Phi^0}' h_{tt}'+\frac{\rho}{\alpha'}e^{2\Phi^0}=0
\end{equation}
$$h_{xx}''- 2{\Phi^0}' h_{xx}'-\frac{\rho}{\alpha'}e^{2\Phi^0}=0 \,$$
\begin{equation}
\label{E29}
h_{tt}''-h_{xx}''+4{\varphi}''+\frac{\rho}{\alpha'}e^{2\Phi^0}=0
\end{equation}
\begin{equation}
\label{E30}
h_{\mu\nu}''-2{\Phi^0}' h_{\mu\nu}' =0 \qquad (\mu \neq \nu ).
\end{equation}
Here primes stand for derivatives with respect to $z$. The term
$H_{\mu\nu}^2$ is asymptotically of order $O(e^{-2\gamma |z|})$ and can be
neglected as a small correction to the leading terms. 
Integrating eqs. (\ref{E28})  we get
\begin{equation}
\label{E31}
h_{tt}' = e^{2\Phi^0} \left(-\frac{\rho }{\alpha'}z+c_t \right)
\end{equation}
\begin{equation}
\label{E32}
h_{xx}'= e^{2\Phi^0 }\left(\frac{\rho }{\alpha'}z+c_x \right)
\end{equation}
and
\begin{equation}
\label{E33}
h_{\mu\nu}'= e^{2\Phi^0 }c_{\mu\nu}
\end{equation}
The constants $c_t$ and $c_x$ are assumed to be independent of $\rho$ and
can be defined by taking the limit $\rho=0$
and comparing with the tree-level solution. In the same way, assuming that
the constants $c_{\mu\nu}$ are $\rho$-independent and comparing (\ref{E33})
with the tree-level solution, we set $c_{\mu\nu}=0$.
Integrating eq. (\ref{E29}) we get
$$
h_{tt}'-h_{xx}'+4{\varphi}'+\frac{\rho}{\alpha'}\int dze^{2\Phi^0}=const.
$$
Adjusting the const to have asymptotically vanishing solution, we obtain
\begin{equation}
\label{E34}
h_{tt}'-h_{xx}'+4\varphi' =
-\frac{\rho}{\alpha'\gamma}\mbox{sgn}ze^{a-\gamma |z|}
\end{equation}
On the other hand, linearizing eq. (\ref{E25}) about the flat-space
solution, we have
\begin{equation}
\label{E35}
h_{tt}''-h_{xx}''-\frac{2\Lambda}{\alpha'} -4{({\Phi^0}')}^2 -
8{\Phi^0}'\varphi'+4{\varphi''}-8{\Phi^0}'(h_{tt}'-h_{xx}') =0.
\end{equation}
Noting that 
$$-\frac{2\Lambda}{\alpha'} -4({\Phi^0}')^2=0$$
is the equation for the vacuum dilaton, equation (\ref{E35}) can be rewritten 
in a form
$$\left(\frac{d}{dz} -2{\Phi^0}' \right)(h_{tt}'-h_{xx}'+4\varphi')=0$$
and solved as
\begin{equation}
\label{E36}
h_{tt}'-h_{xx}'+4\varphi' = ce^{2\Phi^0}
\end{equation}
Comparing (\ref{E34}) and (\ref{E36}), we see that both forms of solution
are equivalent up to exponentially small corrections if we make the
identification $c=-\frac{\rho}{\alpha'\gamma}\mbox{sgn}z$.

{\bf 6.}

Having obtained loop-corrected asymptotics of backgrounds, we can
calculate string-loop-corrected mass of black string. In 
standard gravity interacting with matter, for a class of metrics which
asymptotically sufficiently quickly approach the flat-space metric, the
total energy of a field configuration can be defined in the framework of
canonical approach \cite{rt,fa,gt} \footnote{It should be mentioned that 
in the absence of a preferred asymptotic frame, the notions of ADM energy 
and  mass are nonunique. As usual, we introduce energy as a quantity 
conjugate to variable $t$.}.
 The total energy is defined as the value of the
hamiltonian taken on the shell of zero constraints $\{\Psi\}=0$.
The resulting expression is of the form of 
space integral over the total derivative
\begin{equation}
\label{E37}
E=-\frac{1}{\kappa_D}\int\,d^{D-1} x 
\partial_i (\sqrt{|G_{D-1}|}f^i )|_{\{\Psi\}=0},
\end{equation}
where
$$ f^i =G_{lm,k}(E^{il}E^{km} - E^{ik}E^{lm})\,$$
Here $G_{ik}$ is the spacial $(D-1)$-dimensional part of the metric, 
$G_{D-1} = \det G_{ik}$, and $E^{ik}G_{kl}= \delta^i_l$. For solutions with
$G_{0i}=0$ this formula is simplified because in this case $E^{ik}=G^{ik}$.

 In dilatonic gravity, separating the $\alpha'$ dependence in $a_0 =
N(\alpha')^{-D/2}$ and introducing the D-dimensional gravitation constant
$$ \frac{1}{\kappa_D}=\frac{N}{2}(\alpha')^{-\frac{D-2}{2}},$$
the action (\ref{E22}) is written as
\begin{equation}
\label{E38}
S=-\frac{1}{\kappa_D}\int d^Dy\,\sqrt {|G|}
e^{-2\Phi}\,\left[\hat{R}-4(D\Phi)^2 +4D^2 \Phi -\frac{2\Lambda}{\alpha'}
+ \frac{\rho}{\alpha'}e^{2\Phi} \right].\,
\end{equation}
 For the genus zero part of the  action $(\rho=0)$, for solutions which
sufficiently rapidly approach the vacuum solution $(\eta_{\mu\nu}, \Phi^0)$, 
calculations similar to those in the standard gravity give \cite{mi}
\begin{equation}
\label{E39}
E=-\frac{1}{\kappa_D} \int\,d^{D-1} x \partial_i \left [e^{-2\Phi}
\sqrt{|G_{D-1}|}
(f^i-4G^{ik}\partial_k \varphi )\right ]|_{\{\Psi\}=0},
\end{equation}
Here we again assumed that $G_{0k}=0$. The expression (\ref{E39}) is valid
also for the action (\ref{E38}) containing the one-string-loop correction,
because the latter contributes only to the potential part of the action  
(\ref{E38}).

In the case $D=3$, for solutions sufficiently rapidly approaching the vacuum
configuration, the divergence in the integrand in (\ref{E39}) is 
asymptotically equal to
$$  \left [e^{-2\Phi^0} (h_{xx}'-4\varphi')\right]'.\,$$
Coordinate $x$ asymptotically measures distances along the string 
and $z$ is the transverse coordinate.
Substituting the asymtotic expression for the black string 
solution (\ref{E34}) in (\ref{E39})
we obtain the mass of the black string per unit length
\begin{equation}
\label{E40}
E=-N(\alpha')^{1/2} \left(e^{-2\Phi^0}h_{tt}'-\frac{\rho}{\alpha'\gamma} \right
)|_{z\rightarrow\infty}.
\end{equation}
For the tree-level solution $(\rho=0)$, the expression (\ref{E40}) reproduces
the the mass of the black string calculated in ref. \cite{hh}:
$$
E|_{\rho=0}=\frac{8N}{\sqrt{k}}(1+\lambda)e^{-a}
$$
where we substituted $\alpha'=\frac{1}{k}$.

If string-loop correction is taken into account, $h_{tt}'$ contains a term
linear in $z$, and expression (40) diverges, the divergence being
proportional to $\rho$. This means that, modifying the tree-level bosonic
string action by one-loop corrections, one cannot define finite ADM mass for
the black-string solution. If one defines the energy by subtracting  the
infinite part which is proportional to the mixing parameter $\rho$, one 
again obtains an expression independent of $\rho$.  

{\bf 7.}

It is well known  that the $3D$ black-string solution is dual to the
spherically-symmetric $3D$ black hole solution \cite{btz,bhtz,hhs,hw}. To have 
the dual solution in the standard form, solution (\ref{E2})-(\ref{E4}) is written as
\begin{equation}
\label{E41}
ds^2= -\left(1-\frac{r_{+}^2}{r^2} \right)d{\hat t}^2 +
\left(1-\frac{r_{-}^2}{r^2} \right)d{\hat x}^2 +
{\left(1-\frac{r_{+}^2}{r^2} \right)}^{-1} 
{\left(1-\frac{r_{-}^2}{r^2} \right)}^{-1}\frac{l^2 dr^2}{r^2}
\end{equation}
$$ B_{\hat{x}\hat{t}}=\frac{r_{+}r_{-}}{r^2},\qquad \Phi=
\frac{1}{2}(a-\ln r^2).\,$$
In new variables
$$
t=al(\hat{x}-\hat{t}),\quad \varphi=a(r_{+}^2\hat{t} -r_{-}^2 \hat{x}),\quad
a=(r_{+}^2-r_{-}^2)^{-1/2}
$$
it takes the form
\begin{equation}
\label{E42}
ds^2=-\left(M-\frac{J^2}{4r^2}\right)dt^2+\frac{2}{l}dtd\varphi +
\frac{1}{r^2}d\varphi^2  + \left(\frac{r^2}{l^2}-M+\frac{J^2}{4r^2}\right)^{-1}
dr^2                         
\end{equation}
$$ B_{\varphi t} =-\frac{J}{2r^2},\qquad \Phi=\frac{1}{2}(a-\ln r^2).\,$$
where
$$ M=\frac{r_{+}^2+r_{-}^2}{l^2},\qquad J=\frac{2r_{+}r_{-}}{l}$$
Dual (with respect to the variable $\varphi$) transformation of the fields 
(\ref{E42}) gives the solution 
\begin{equation}
\label{E43}
d{\tilde{s}}^2=-\left(M-\frac{r^2}{l^2}\right)dt^2+Jdtd\varphi +
r^2 d\varphi^2 + \left(\frac{r^2}{l^2}-M+\frac{J^2}{4r^2}\right)^{-1}dr^2                         
\end{equation}
$$ B_{\varphi t} =-\frac{r^2}{l^2},\qquad \Phi=0.\,$$
The metric $d{\tilde{s}}^2$ is the black-hole solution in $3D$ Einstein
gravity \cite{btz,bhtz,hhs,hw}. The fields (\ref{E43}) are solutions of equations of motion 
for the action (\ref{E5}) with the cosmological constant
$\Lambda=-\frac{2\alpha'}{l^2}$. Duality transformation \cite{b} leaves the 
form of equations of motion unchanged.

Let us consider duality transformations in the theory with the loop-corrected 
action (\ref{E23}). Requiring that equations of motion (\ref{E24})-(\ref{E26})
do not change their functional form under duality transformations 
(cf. \cite{b})
$\varphi^i=\varphi(\tilde{\varphi})+\rho\delta\varphi(\tilde{\varphi})$, we 
have
$$
\beta^i (\varphi)+\rho\delta\beta^i (\varphi)|_{\varphi^i=
\varphi(\tilde{\varphi})+\rho\delta\varphi(\tilde{\varphi})}=
\beta^i (\tilde{\varphi})+\rho\delta\beta^i (\tilde{\varphi}).
$$
Keeping terms linear in $\rho$, we obtain
$$\delta\beta+\delta\varphi^j \frac{\partial \beta^i}{\partial\varphi^j}|_
{\varphi^i = \varphi^i (\tilde{\varphi})} =\delta\beta^i(\tilde{\varphi})
$$
The number of equations on the functions $\delta\varphi^i$ is equal the number
of the functions $\delta\varphi^i$. Finding the functions $\delta\varphi^i$,          
we obtain the duality transformations which leave the functional form of the 
loop-corrected $\beta$-equations unchanged.

{\bf 8. Conclusions and discussion.}

In this paper, starting from string-loop-corrected renormalized EA, we
calculated one-string-loop corrections to black-string backgrounds which,
on one hand, are obtained from the gauged WZW model, and, on the other
hand, are solutions to $O(\alpha')\quad \beta$-function equations derived
from tree-level EA. Although final calculations were performed for an
example of ${SL(2,R)\times R}/R $ WZW model, our discussion was quite general:
all the expressions can be written in D-dimensional form and applied
to the case of a general ${SL(2,R)\times R^{N}}/R$ model. It was found that
backgrounds acquire corrections of order $ \rho \int [d^2 \tau] $,
where $\rho$ is parameter accounting for an admixture of genus-one 
contribution to the tree-level part of EA. 

From (15) it follows that for all ${SL(2,R)\times R^N}/R$ models and, in
particular, for the $3D$ black-string solution,  the integral over
the moduli is exponentially divergent. This divergence is the well-known
tachyonic divergence in bosonic string theory, which 
is absent in superstring theory. In paper \cite{bi} this problem was 
discussed in the framework of fermionic string theory 
and it was argued that in this theory there appear 
no terms in EA which could give divergent corrections to tree-level result.

Using the one-string-loop-corrected backgrounds obtained by solving the
$\beta$-equations, we calculated the ADM mass of the black string. It
appeared that the result is divergent, the divergence being proportional to  
the mixing parameter $\rho$. Redefining the energy (mass) by
subtracting the infinite part, one obtains an expression independent of $\rho$.
Thus, in the present case, the conjecture of ref. \cite{emn} about the
imaginary string-loop corrections to the tree-level mass does not work.

Although our calculations were restricted to one-loop contributions, higher-
order corrections can be discussed as well. Loop-corrected EA has the 
following structure:
$$\hat{Z} = \hat{Z}_{0} + \hat{Z}_{1} + \hat{Z}_{2} + \ldots =$$
\begin{equation}
a_0\,\int d^D x \sqrt{G} e^{-\Phi} (-2\Lambda + \alpha' R + \ldots) +
a_1\,\int d^D x \sqrt{G} (1 + \ldots) + \,
\end{equation}
$$a_2\,\int d^D x \sqrt{G} e^{\Phi}(1 + \ldots) + \ldots$$
(here $a_i$ include integrals over the moduli). It is seen that higher-%
genus corrections are accompanied by the factors $e^{\frac{\chi\Phi}{2}}$
and are exponentially suppressed at spatial infinity for the black-string
solution in question. Thus, in any finite order in string-loop expansion, 
corrections from higher topologies will not  contribute to the ADM mass.

For the $SL(2,R)/R$ model, loop corrections can be calculated in the same
way as above. However, in this case, because of 2D relation $R_{\mu\nu} = 
\frac{1}{2} G_{\mu\nu} R, \quad \beta$-equations are much simpler and can be 
easily solved exactly. In notations of \cite{msw} we have
$$ \Phi = \frac{Q \eta}{2} $$
$$ G_{\mu\nu} = diag [-g(\eta), g(\eta)^{-1}] $$
where $g(\eta)$ is now solution of loop-corrected equation
\begin{equation}
g'' = Qg' + \frac{\rho}{\alpha'} e^{Q\eta}
\end{equation}
of the form
$$ g(\eta) = 1 + ae^{Q\eta} + \frac{\rho\eta}{\alpha' Q} e^{Q\eta}.$$
Note that in this case, as in $3D$ theory there appear the term
$O(\eta e^{Q\eta})$.
The relation  $Q^2=|\Lambda|$ does not recieve loop corrections
and is the same as at the tree level. Since in 2D theory there is no 
tachyon in the spectrum, integration measure over the moduli contains 
no exponential factors.

{\bf Acknowledgements.}

I would like to thank V.Belokurov, R.Metsaev and I.Tyutin for useful
discussions.

\end{document}